\documentclass{llncs}

\usepackage{llncsdoc}
\usepackage{graphicx}
\usepackage{algorithm}
\usepackage{algorithmic}

\newcommand{\keywords}[1]{\par\addvspace\baselineskip
\noindent\keywordname\enspace\ignorespaces#1}

\begin {document}
\title{Core First Unit Propagation}
\titlerunning{Core First Unit Propagation}

\author{Jingchao Chen}
\institute{School of Informatics, Donghua University \\
2999 North Renmin Road, Songjiang District, Shanghai 201620, P. R.
China \email{chen-jc@dhu.edu.cn}}

\maketitle

\begin{abstract}

Unit propagation (which is called also Boolean Constraint Propagation) has been an important component of every modern CDCL SAT solver since the CDCL
solver was developed. In general, unit propagation is implemented by scanning sequentially every clause over a linear watch-list.
This paper presents a new unit propagation technique called core first unit propagation. The main idea is to prefer core clauses over
  non-core ones during unit propagation, trying to generate a shorter learnt clause. Here, the core clause is defined
  as one with literal block distance less than or equal to 7.
  Empirical results show that core first unit propagation improves the performance of the
winner of the SAT Competition 2018, MapleLCMDistChronoBT.


\end{abstract}

\section{Introduction}

Since the GRASP solver was envisioned in 1996 \cite {GRASP}, Conflict-Driven Clause Learning (CDCL) SAT solving has been
achieved great success in many fields. Unit propagation (which is called also Boolean Constraint Propagation) is not only
an important component of every modern CDCL SAT solver, but also an important one of some proof checkers \cite {GRAT}.
To our best knowledge, so far this component has not been studied yet.  This paper focuses on this problem.

A CDCL SAT solver works on a CNF (Conjunctive Normal Form) formula, which is defined as a finite conjunction of clauses, and also can
be denoted by a finite set of clauses. A clause is a disjunction of literals, also written as a set of literals, which is either a variable or the negation of a variable. A clause is said to be a unit clause
if it consists only of literals
assigned to value 0 (false) and one unassigned literal.
BCP (Boolean Constraint Propagation) fixes the unassigned literal in a unit clause to the value 1 (true) to
satisfy that clause. This variable assignment is referred to as an implication.
BCP carries out repeatedly the identification of unit clauses and the creation of the associated
implications until either no more implications are found or a conflict (empty clause) is produced.

It is generally accepted that BCP is implemented
by scanning sequentially every clause over a linear watch-list. This implementation is called a standard BCP. By our empirical observation,
we found that the standard BCP implementation is not efficient in some cases. Therefore, we decided to propose a new unit propagation technique
called core first unit propagation.
The basic idea of this technique is to prefer core clauses over
  non-core ones during unit propagation, trying to generate a shorter learnt clause.
  Here£¬ the core clause is defined
  as one with literal block distance less than or equal to 7. This definition is consistent with that of Ref. \cite {LBD3tier}.
  Empirical results show that core first unit propagation improves the performance of the
winner of the SAT Competition 2018, MapleLCMDistChronoBT \cite{CBT1,CBT2}.

\section{Core First Unit Propagation}

The idea of CFUP ( Core First Unit Propagation) is to classify clauses as core or non core, and prefer core clauses over
  non-core ones during unit propagation.
  A clause is core if it is a learnt clause and its LBD (Literal Block Distance) value is less than 7.
  LBD is defined as the number of decision variables in a clause \cite {LBD}.
  Our core concept corresponds to the concept of non local in the CoMiniSatPS solver that classifies learnt clauses into
 three categories \cite {LBD3tier}. References \cite {GRAT,TreeRat} have similar concepts. But they are different from the
 core concept used in this paper. In~\cite {GRAT,TreeRat}, core clauses refer to marked or visited ones, and have nothing to do with LBD.

 Our CFUP uses a single watchlist, not two separate watchlists. We implement to select core clauses first by moving core clauses
ahead of non-core clauses during unit propagation. When watchlists are built initially, core clauses are not in front of non-core clauses.
Like the standard BCP, the goal of CFUP is to search for all unit clauses. This can be done by repeating the following process until
either no more implications are found or a conflict (empty clause) is produced:
Remove the first unvisited literal $l$ from $T$; get new implications from clauses watched by $l$; and add the new implications to $T$,
where $T$ is a trail stack of decision literals and implications. The core priority strategies of CFUP embodies in the update of watchlists.
Algorithm~\ref{alg1} shows CFUP.

The pseudo-code of CFUP shown in Algorithm~\ref{alg1} assumes that a full literal watch scheme (a full occurrence list of all clauses) is used,
If using a two literal watch scheme \cite {SATO}, The statement ``Append $W[l]-C$ to the end of $C$ " in Algorithm~\ref{alg1} can be modified as follows.

\begin{algorithm}
$W[l]$: set of clauses watched by literal $l$

\begin{algorithmic}
\STATE $D :=\emptyset $
\FOR{ $k=0$ to end index of $W[l]$ }
    \IF {$W[l][k]$ has more than two unassigned literals}
    \STATE $D := D \cup \{W[l][k]\}$
    \STATE $W[\overline{s}] = D \cup \{W[l][k]\}$\\
     where $s$ is unwatched and unassigned literal
    \ENDIF
\ENDFOR
\STATE Append $W[l]-C-D$ to the end of $C$
\end{algorithmic}
\end{algorithm}

\begin{algorithm}[tb]
\caption{CFUP( ): Core First Unit Propagation}
\label{alg1}
\emph{T}: trail stack of decisions and implications\\
$W[l]$: set of clauses watched by literal $l$

\begin{algorithmic}
\STATE $\beta := null $
\FOR {$q:=$ index of 1st unvisited literal in $T$  to $T$.size}
   \STATE $l := T[q]$ \\
    $C :=\emptyset$, where $C$ is used to store core clauses
    \FOR { $k=0$ to $W[l]$.size }
         \IF {$W[l][k]$ is unit}
             \STATE  $u:=$ the unassigned literal of $W[l][k]$ \\
             Push $u$ to the end of $T$ \\
             \IF {$W[l][k]$ is core clause}
                \STATE $C := C \cup \{W[l][k]\}$
             \ELSE
                 \IF { $W[l][k]$ is falsified}
                 \STATE  $\beta := W[l][k]$  {\bf break}
                 \ENDIF
             \ENDIF
          \ENDIF
          \STATE Append $W[l]-C$ to the end of $C$ \\
           $W[l] := C$ \\
           {\bf if} {$\beta \neq null $} {\bf then return} $\beta$
    \ENDFOR
\ENDFOR
\STATE {\bf return} $null$
\end{algorithmic}
\end{algorithm}

\begin{algorithm}[tb]
\caption{CDCL(): Conflict-Driven Clause Learning}
\label{alg2}
%\textbf{Input}: Your algorithm's input\\
%\textbf{Parameter}: Optional list of parameters\\
%\textbf{Output}: Your algorithm's output
\emph{T}: trail stack of decisions and implications\\
\textbf{$\gamma$}: a learnt clause
%\begin{algorithmic}[1] %[1] enables line numbers
\begin{algorithmic}
\WHILE{not all variables assigned}
  \IF {$No\_of\_conflict > \theta$}
  \STATE $ conflict\_cls := $ \textbf{BCP}()
  \ELSE
  \STATE $ conflict\_cls := $ \textbf{CFUP}()
  \ENDIF
  \IF {$conflicting\_cls \neq null$}
    \STATE $No\_of\_conflict := No\_of\_conflict +1 $
    \STATE $(1uip, \gamma) :=$ \textbf{ConflictAnalysis}$(conflicting\_cls)$
     \IF {$\gamma = \emptyset$}
     \STATE {\bf return} UNSAT
     \ENDIF
     \STATE Push $1uip$ to $T$
     \STATE \textbf{Backtrack}(current decision level-1)
  \ELSE
  \STATE Decide and push the decision to $T$
  \ENDIF
\ENDWHILE
\STATE \textbf{return} SAT
\end{algorithmic}
\end{algorithm}

%\STATE Let $t=0$.
%\STATE Do some action.

Removing the statement ``$C := C \cup \{W[l][k]\}$" in CFUP yields a standard BCP. In the real implementation, we do not use a list to store
core clauses during unit propagation. instead of it, we do it by swapping two elements in $W[l]$. In details, let $W[l][0..m]$ and $W[l][m+1..k-1]$
be core and non core clause zone, respectively. if $W[l][k]$ is a core clause, we swap $W[l][k]$ and $W[l][m+1]$. Otherwise, we do nothing.
A general CDCL solver has two watchlists: binary and non binary. We adopt the core priority strategy only on a non-binary watchlist.

%\clearpage

By our empirical observation,  adopting always the core priority strategy is not good choice. A better policy is that when
the number of conflicts is less than $2\times 10^6$, \textbf{CFUP} is called, Otherwise, \textbf{BCP} is called. The high-level algorithm CDCL combining
\textbf{CFUP} and \textbf{BCP} are shown in Algorithm~\ref{alg2}.

 CDCL given in Algorithm~\ref{alg2} uses  a
loop to reach a status where either all the variables are assigned (SAT) or an
empty clause is derived (UNSAT). Inside the loop, based on whether the number of conflicts is greater
than $\theta$, it decides to invoke either \textbf{CFUP} or \textbf{BCP}. Here \textbf{BCP} is considered a
unit propagation without any priority strategy.
If there is a conflict, \textbf{CFUP} or \textbf{BCP} returns a falsified conflicting clause. Otherwise, a new
decision is taken and pushed to the trail stack. Conflict analysis learns a new 1UIP clause $\gamma$.
CDCL asserts the unassigned 1UIP literal and pushes it to the trail stack.

\section{Empirical evaluation}

%\hyphenation{seconds}

All experiments were conducted under the following platform: Intel core i5-4590 CPU with speed of
3.3 GHz. The timeout for each solving was set to 5000 seconds.
We have added \textbf{CFUP} to MapleLCMDistChronoBT \cite{CBT1,CBT2}, which was the
winner of the main track in the SAT Competition 2018 \cite {SAT18}.

  Table~\ref{Tab} shows briefly the runtime and solved instances of the default Maple-LCMDistChronoBT
  vs. the best configuration in CFUP mode, $ \theta = 2 \times 10^6 $, as well as two vicinity configurations
$ \theta = 10^6 $ and $ \theta = 3 \times 10^6 $. As seen in Table~\ref{Tab}, $ \theta = 2 \times 10^6 $ outperforms
the default MapleLCMDistChronoBT in terms of both the number of solved instances and the
runtime. It solves 5 more instances and is faster by 5682 seconds.
The number of core clauses increases with the increase of the number of conflicts. When the number of core clauses is large, CFUP is identical to BCP. Compared with BCP,
the cost of CFUP is higher than that of BCP. So $\theta$ should not be set to very large.
It is easy to see that CFUP
has a certain extent advantage on satisfiable instances in some configurations.

\begin{table}
\caption{
   Runtime (in seconds) and solved instances of MapleLCMDistChronoBT on SAT competition 2018 instances
} \label{Tab}
\begin{center}

\setlength\tabcolsep{4pt}
%\begin{tabular}{p{4.3cm}|c|c|c|c|c}
\begin{tabular}{l|l|c|c|c|c}
\hline  \hline
\multicolumn{2}{c|}{                   } & Base & $ \theta = 10^6 $ &  $ \theta = 2 \times 10^6 $ & $ \theta = 3 \times 10^6 $\\
\hline
                             &  Solved &   138    &   134     & 142    &  141 \\
\raisebox{1.1ex}[0pt]{SAT}   &  Time   &  99104   &   78397   & 91136  & 95723 \\
\hline
                             &  Solved &   102    &   102     & 103    &  103  \\
\raisebox{1.1ex}[0pt]{UNSAT} & Time    & 66845    &   70338   & 69131  &  72962\\
\hline
                             &  Solved &   240    &   236     & 245    &  244 \\
\raisebox{1.1ex}[0pt]{ALL}   &  Time   & 165949   &   148735  & 160267 &  168685 \\

\hline
\end{tabular}
\end{center}
\end{table}

\begin{figure}
\centering
\includegraphics[height=6.8cm]{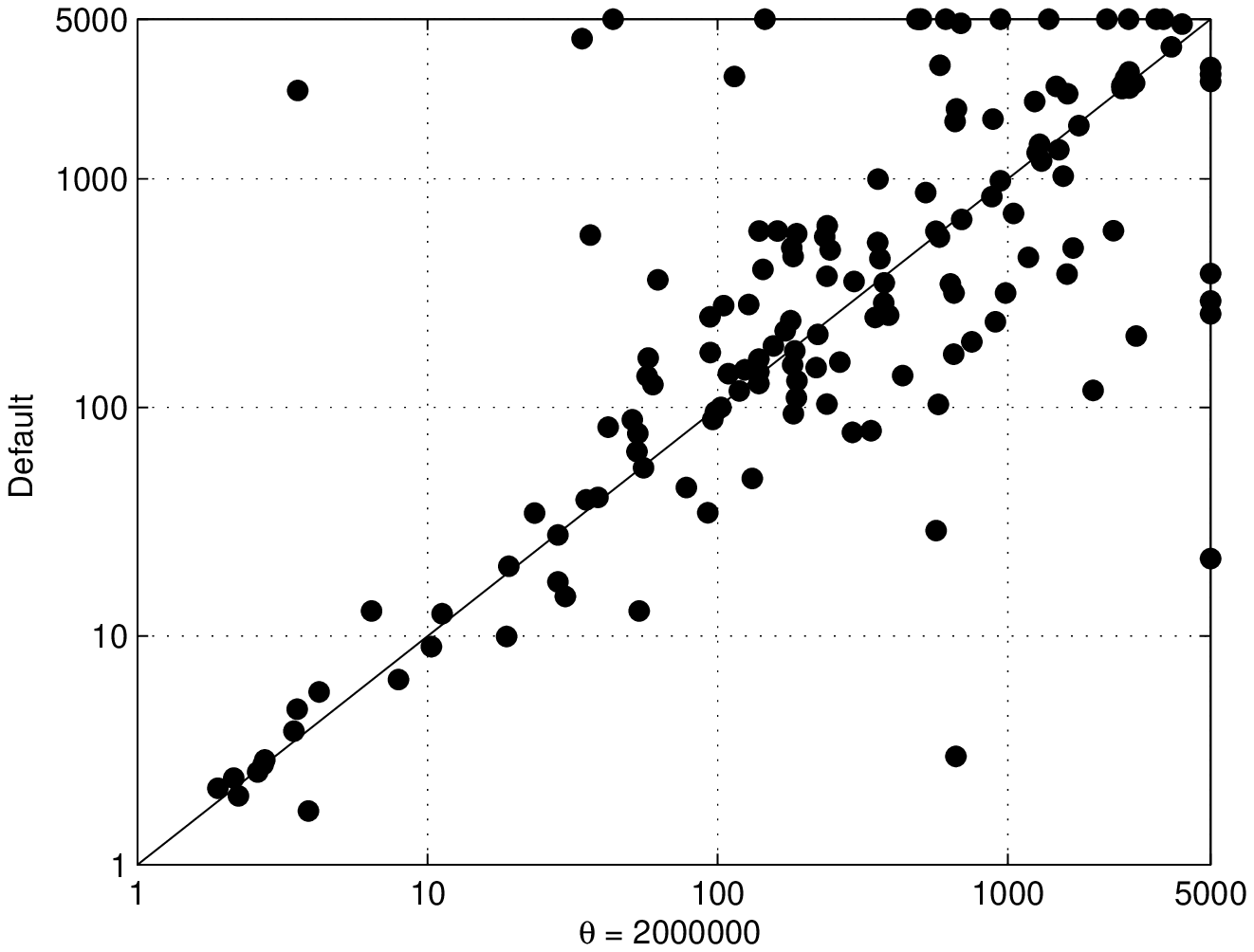}
\caption{MapleLCMDistChronoBT on SAT} \label{scaterFig1}
\end{figure}

\begin{figure}
\centering
\includegraphics[height=6.8cm]{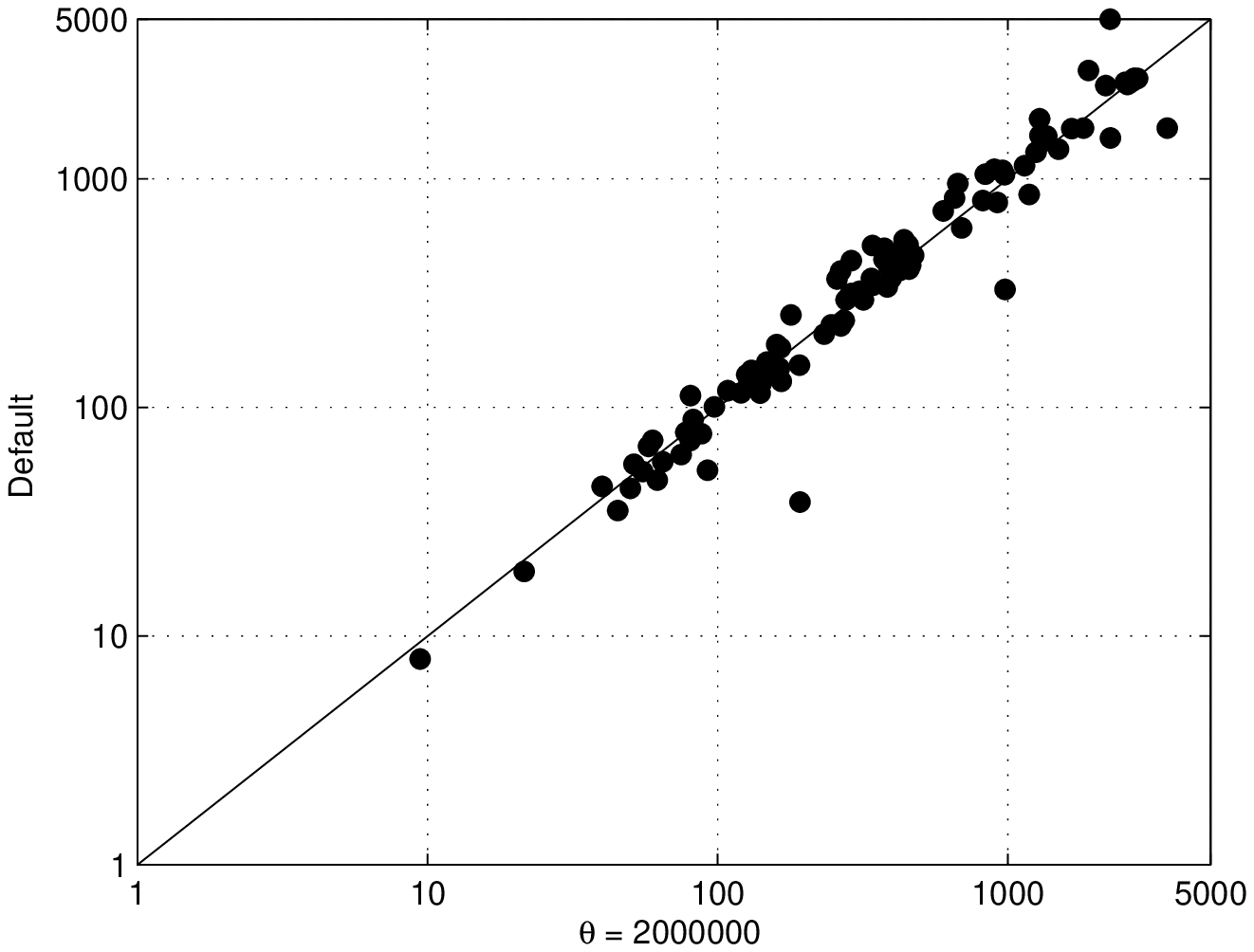}
\caption{MapleLCMDistChronoBT on UNSAT} \label{scaterFig2}
\end{figure}

Figures \ref{scaterFig1} and \ref{scaterFig2} shows a log-log scatter plot comparing the running times
of the default MapleLCMDistChronoBT vs.
the overall winner $ \theta = 2 \times 10^6 $ on satisfiable and unsatisfiable instances,
respectively.
Each point corresponds to a given instance. A point at line $y=5000$ (resp., $x=5000$) means that the instances on that point were not
solved by default version (resp., $ \theta = 2 \times 10^6 $). As shown in Figure~\ref{scaterFig1}, the points
that appear over the diagonal are more than ones below the diagonal. Figure \ref{scaterFig1} shows that in many cases, $ \theta = 2 \times 10^6 $) is faster than the default
configuration. Among the instances given in Figure \ref{scaterFig1}, the unsolved instances of the default configuration
 and the best configuration are 12 and 8, respectively. That is, the best configuration solves  4 more satisfiable instances
 than the default configuration. Figure \ref{scaterFig2} demonstrates that although in almost all the cases the speed of the best configuration
 is the same as that of the default configuration, the best configuration solves 1 more unsatisfiable instance than the default.

%\clearpage

\section{Conclusions}

Implementing CFUP is a trivial task. It can be done by making a little modification to BCP of the solver.
We have added CFUP into the main track winner of the SAT Competition 2018, MapleLCMDistChronoBT.  Empirical results
show that CFUP improves the overall performance of the solver in some configurations. In theory, when analyzing a conflicting clause,
using short LBD clauses should be more beneficial than using long LBD clauses. That is, replacing completely the standard BCP with
CFUP should be the best choice. However, in fact, combining CFUP and the standard BCP is a good choice.  Its reason is well worth studying
in future.

\begingroup
\hyphenation{Jarvisalo}
\hyphenation{proofs}
\hyphenation{Laurent}

%\raggedleft
%\raggedright
%\hfill

\bibliographystyle{named}
\bibliography{coreFirstIJCAI19}

\endgroup
\end{document}